\documentstyle[prl,aps,epsf]{revtex}

\begin{document}
\draft
\twocolumn[\hsize\textwidth\columnwidth\hsize\csname@twocolumnfalse\endcsname

\title{Inversion Symmetry and Critical Exponents of
Dissipating Waves in the Sandpile Model}
\author{
Chin-Kun~Hu$^{1}$,
E.V.~Ivashkevich$^{1,2}$,
Chai-Yu~Lin$^{1}$ and
V.B.~Priezzhev$^{2}$}
\address{
$^{1}$Institute of Physics, Academia Sinica,
Nankang, Taipei 11529, Taiwan\\
$^{2}$Bogoliubov Laboratory of Theoretical Physics,
Joint Institute for Nuclear Research, Dubna 141980, Russia}
\date{\today}
\maketitle

\begin{abstract}
Statistics of waves of topplings in the sandpile
model is analyzed both analytically and numerically.
It is shown that the probability distribution of
dissipating waves of topplings that touch the boundary
of the system obeys
the power-law with critical exponent
$5/8$. This exponent is not independent and is related
to the well-known exponent of the probability distribution
of last waves $11/8$ by exact inversion symmetry
$s\to 1/s$. Probability distribution of those dissipating
waves that are also last in an avalanche is invariant under
the inversion transformation and has asymptotic behavior
$s^{-1}$.
Our extensive numerical simulations not only
support these predictions, but also indicate that inversion
symmetry is also useful for the analysis of the two-wave
probability distributions.
\end{abstract}
]

The concept of self-organized criticality (SOC) introduced
by Bak, Tang and Wiesenfeld \cite{bak-87} is considered to be
the underlying cause of a variety of critical phenomena
involving dissipative, nonlinear transport in open systems.
In these phenomena a system with infinite number of degrees
of freedom, when driven by external random force, evolves
stochastically into a certain critical state on which it
exhibits properties similar to that of a second order phase
transitions. This critical state is characterized by strong
fluctuations of the order parameter with power-law decay of
correlation functions and absence of any characteristic
length or time scale. However, unlike second order phase
transitions, in the SOC phenomena the criticality emerges
automatically without fine tuning of any parameters like
temperature or pressure. Kolmogorov theory of isotropic and
homogeneous turbulence \cite{kolmogorov-41} and the theory
of fractal growth proposed by Kardar, Parisi and Zhang
\cite{kardar-86} are probably the most important physical
examples of SOC.

To illustrate the basic ideas of SOC Bak, Tang and
Wiesenfeld introduced a cellular automaton now commonly
known as sandpile model because of the crude analogy
between its dynamical rules and the way sand topples when
building a real pile of sand. One natural formulation of
the sandpile model is given in terms of integer height
variables $z_{i}$ at each site of a planar square lattice
${\cal L}$. In a stable configuration the height $z_{i}$ at
any site $i\in{\cal L}$ takes values $1,2,3$ or $4$.
Particles are added at randomly chosen sites and the
addition of a particle increases the height at that site by
one. If this height exceeds the critical value $z_c=4$,
then the site topples, and on toppling its height decreases
by $4$ and the heights at each of its neighbors $j$
increases by $1$. These neighboring sites may become
unstable in their turn and the toppling process continues
causing an avalanche.

Extensive numerical simulations of this simple model
verified that it captures all the essential features of SOC
\cite{NumSym}. Another reason why a lot of attention has
been attracted to this model during last years is that the
problem admits a purely analytic treatment.
Dhar \cite{dhar-90} discovered the Abelian structure
of the sandpile dynamics and then Dhar
and Majumdar \cite{majumdar-91} proved that
recurrent sandpile configurations are in one-to-one
correspondence with the spanning trees on the lattice.
This result gave a key to exact calculation of height
probabilities in the bulk of the lattice
\cite{majumdar-92,priezzhev-94} and height-height
correlation functions at the boundary \cite{ivashkev-94a}.

All these advances notwithstanding, the derivation of the
probability distribution of avalanches still remains an
important open problem.  Recent numerical simulations
\cite{demenech-98} indicate that this
probability distribution is actually multi-fractal. The
avalanches propagating up to the boundary and, hence,
dissipating particles there seem to form very special
subclass of avalanches which obey pure scaling.  This is
the reason why the analysis of dissipating events becomes
of especial importance.  It is the aim of the Letter to
clarify the issue and to study the probability
distributions of some classes of dissipating events.

The spanning trees representation turned out to be very
useful not only in the study of static height-height
correlations but also in the analysis of the avalanche
process.  Namely, every avalanche may be represented as a
sequence of more elementary events, so-called {\it waves of
topplings}. These can be
organized as follows \cite{ivashkev-94b}: if the site $i$
to which a grain was added becomes unstable, topple it once
and then topple all other sites of the lattice that become
unstable, keeping the initial site $i$ from second
toppling. The set of sites toppled thus far is called {\it
the first wave of topplings}. After the first wave is
completed the site $i$ is allowed to topple the second
time, not permitting it to topple again until {\it the
second wave of topplings} is finished. The process
continues until the site $i$ becomes stable and the
avalanche stops.

Waves of topplings being more elementary events than
avalanches have also much simpler properties. First,
each site involved into a wave topples exactly once in
that wave. As a result, the height profile of the system
after a wave is exactly the same as before the wave except
at sites on the single closed boundary of the wave which
separates sites that toppled in that wave from those that
did not. Just inside the wave boundary a trough (relative
to the previous heights) appears , whereas outside the
boundary a hill appears where sand was moved out of the
wave boundary. The avalanche can stop only if the site
where avalanche was initiated has at least one neighboring
site not belonging to the last wave and therefore the very
last wave in the avalanche should touch the origin of the
avalanche. All the waves are individually compact,
so, we shall characterize the waves of topplings only by
their area.

The  most important of geometrical properties of waves is
the one-to-one correspondence between waves and two-component
spanning trees where one component corresponds to the wave
itself and the other to the rest of the lattice.

The expected number of different waves of topplings is related to
the Green function of the Laplacian operator and was found
to be \cite{ivashkev-94a}
\begin{equation}
W_{all}(s)~{\rm d}s\sim\frac{{\rm d}s}{s}.
\label{AllWaves}
\end{equation}
Note that the distribution
function ${\rm Prob}[s>s_0]$ of all waves is logarithmically
divergent both on the ultraviolet cutoff $a$ (step of the
lattice) and infrared cutoff $L$ (lattice size).

\begin{figure}[t]
\epsfxsize=70mm
\epsffile[30 -20 536 470]{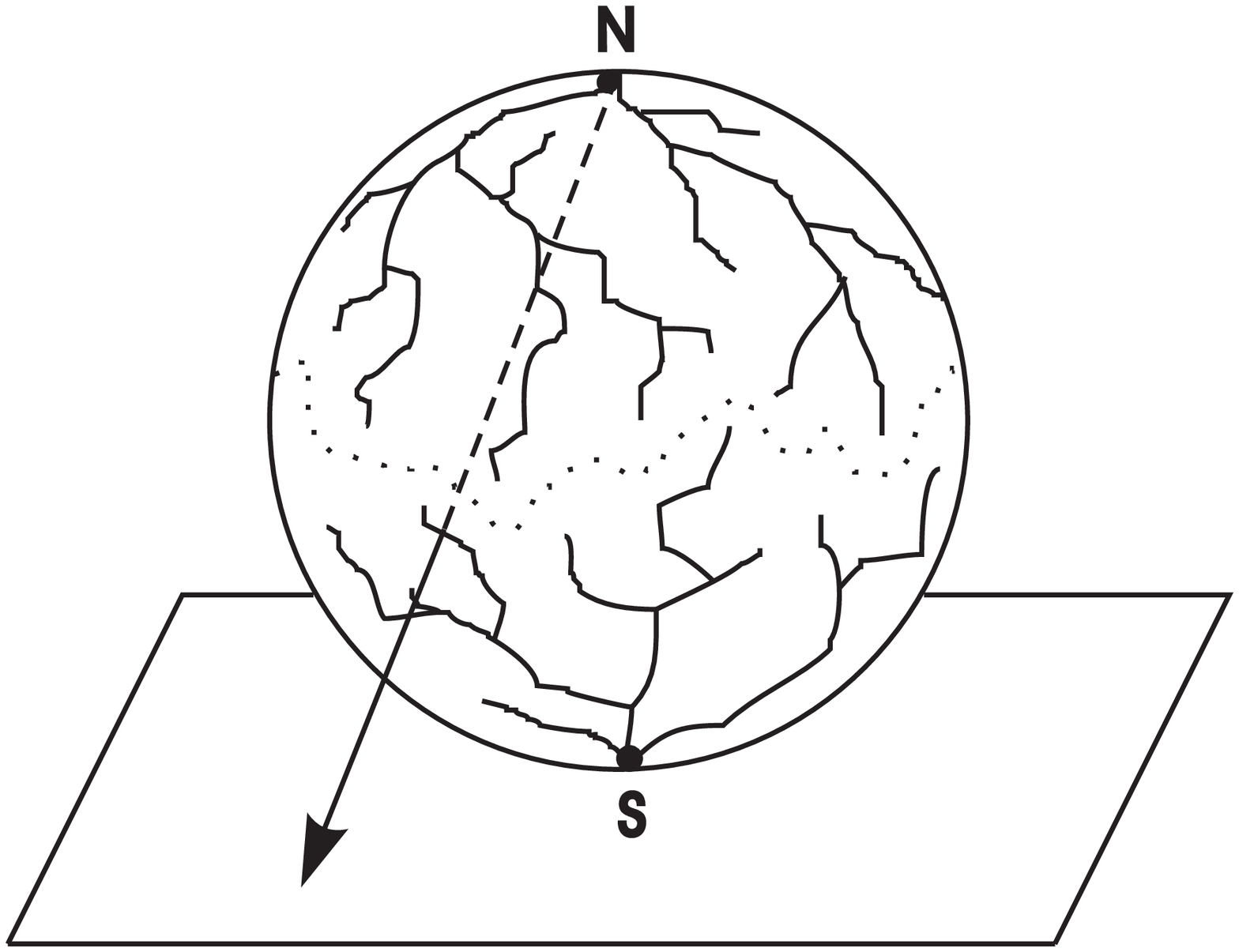}
\caption{Stereographic projection from plane onto sphere
transforms the spanning tree so that one its component
is attached to the south pole of the sphere, and another
to the north.}
\label{fig1}
\end{figure}

This probability distribution is obviously invariant under
the inversion transformation $r\to R^2/r$, where
$R=\sqrt{aL}$ is called an inversion radius.  The meaning
of this transformation can easily been understood as
follows. Let us consider stereographic projection from the
plane to the sphere as shown on Fig.\ref{fig1}.
The boundary of the system in such a projection is mapped
to the north pole while the point where a wave of topplings
was initiated to the south pole.  Two-component spanning
tree on the plane that represents the height configuration
after a given wave of topplings will be transformed into
the similar tree on the sphere.  Its one
component is attached to north pole of the sphere and the
other component to the south pole. Both poles of the sphere
are formally equivalent and inversion transformation
interchanges them.

Inversion transformation is only one of the family of
conformal symmetries. Inversion, however, is of special
importance for us since it provide us with
information about the distribution of dissipating waves of
topplings, i.e. those waves that touch the boundary of the
system and dissipate particles there.
\begin{figure}
\epsfxsize=70mm
\vbox to4.3in{\rule{0pt}{4.3in}}
\includegraphics{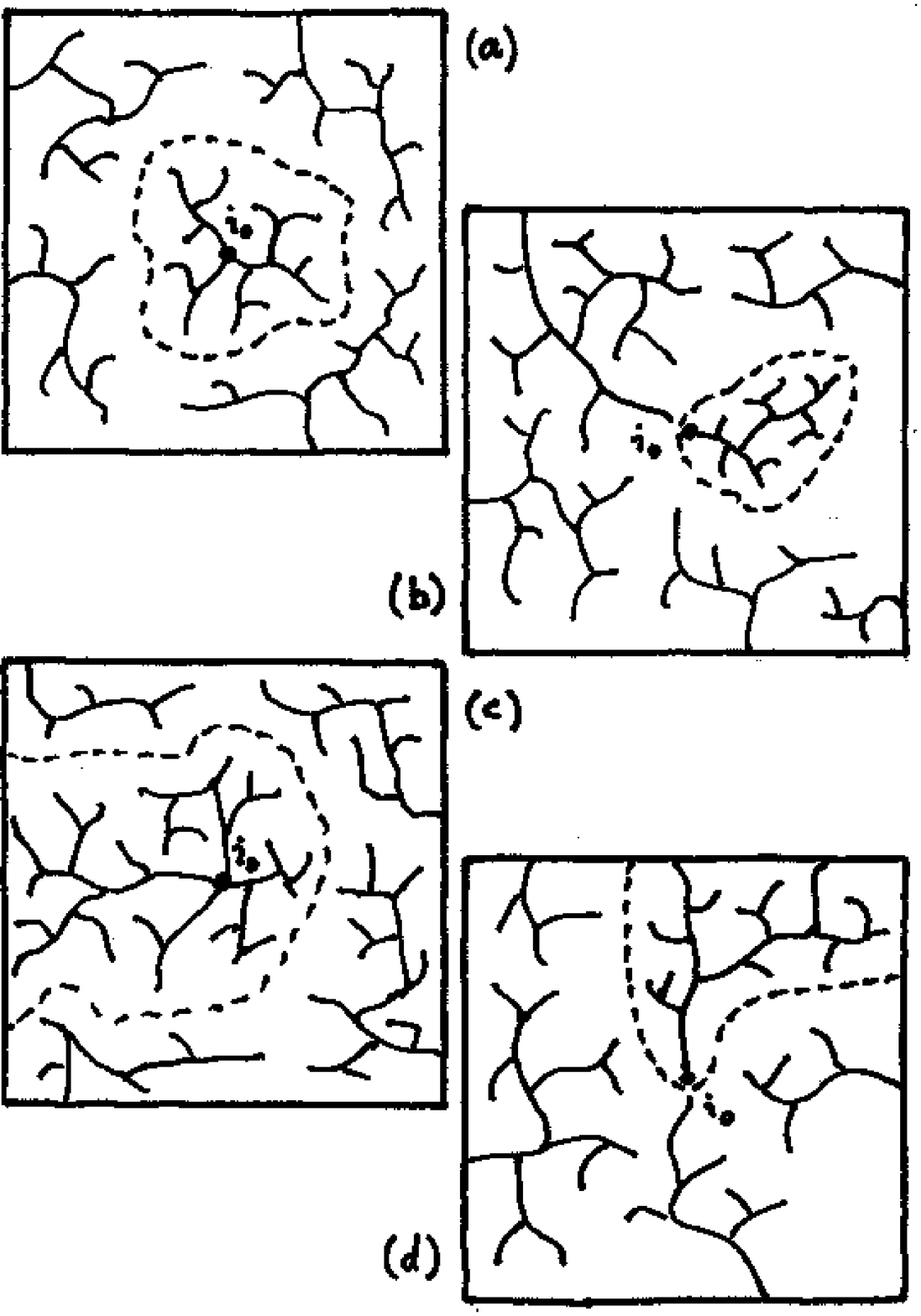}
\caption{\label{fig2}
Two component spanning trees  corresponding to: (a) general,
(b) last, (c) dissipating and (d) dissipating last waves.}
\end{figure}

Indeed, let us consider all waves that, being initiated
somewhere in the bulk of the system, propagate
to the boundary and dissipate particles there.
These are represented by the two-component spanning tress such
that one component touches
the boundary (Fig.2(c)) or the north pole of the sphere.
Then, after inversion transformation this set of
dissipating waves will be transformed to the set of waves
that touch the south pole of the sphere.
This is nothing but characteristic feature of the last wave
of topplings.
The probability distribution of last waves is known exactly
\cite{dhar-94}
\begin{equation}
W_{last}(s)~{\rm d}s\sim
\left(\frac{a^2}{s}\right)^{3/8}
\frac{{\rm d}s}{s}.
\end{equation}
(The ultraviolet cutoff $a$ here appears because the
probability is dimensionless.) Now, since
the inversion transforms last waves
into dissipating ones
and vice versa, we conclude that after substitution $r\to
R^2/r$ the probability distribution
of last waves gets the probability distribution of
dissipating waves.  In this
way, we obtain
\begin{equation}
W_{diss}(s)~{\rm d}s\sim
\left(\frac{s}{L^2}\right)^{3/8}
\frac{{\rm d}s}{s}.
\end{equation}
Then, if we consider only those dissipating waves that are also last
in an avalanche then we can notice that such waves touch both the
boundary of the system and the site where avalanche was initiated.
So, their distribution should be invariant under inversion
transformation, $r\to R^2/r$. The only such a distribution is
\begin{equation}
W_{diss~\!\!\&~\!\!last}(s)~{\rm d}s\sim
\left(\frac{a}{L}\right)^{3/4}\frac{{\rm d}s}{s}.
\end{equation}
Eq.(4) coincides with Eq.(\ref{AllWaves}) up to a prefactor
$\left(a/L\right)^{3/4}$.  The origin of the prefactor is
the following. The conformal field theory which underlies the model
of dense polymers (spanning trees) has central charge
$c=-2$. Saleur and Duplantier
\cite{saleur-87} proved that the dimension of the operator
$\phi_k(\bf{r})$ that corresponds to the insertion of the
polymer star with $k$ legs into the site $\bf{r}$ of the
plane is equal to $\Delta_k=(k^2-1)/4$.
The polymer star is formed by
$k$ different branching polymers joined
at the site $\bf{r}$ and connecting the site with the
boundary of the system. We can note now, that the spanning
tree corresponding to a dissipating last wave form a
two-leg polymer star and, hence, corresponds to the
operator $\phi_2$ with dimension $3/4$
i.e. the average of $\phi_2$
should vary with the size of the system as
$(a/L)^{3/4}$.

\begin{figure}
\epsfxsize=70mm
\vbox to2.9in{\rule{0pt}{2.9in}}
\includegraphics{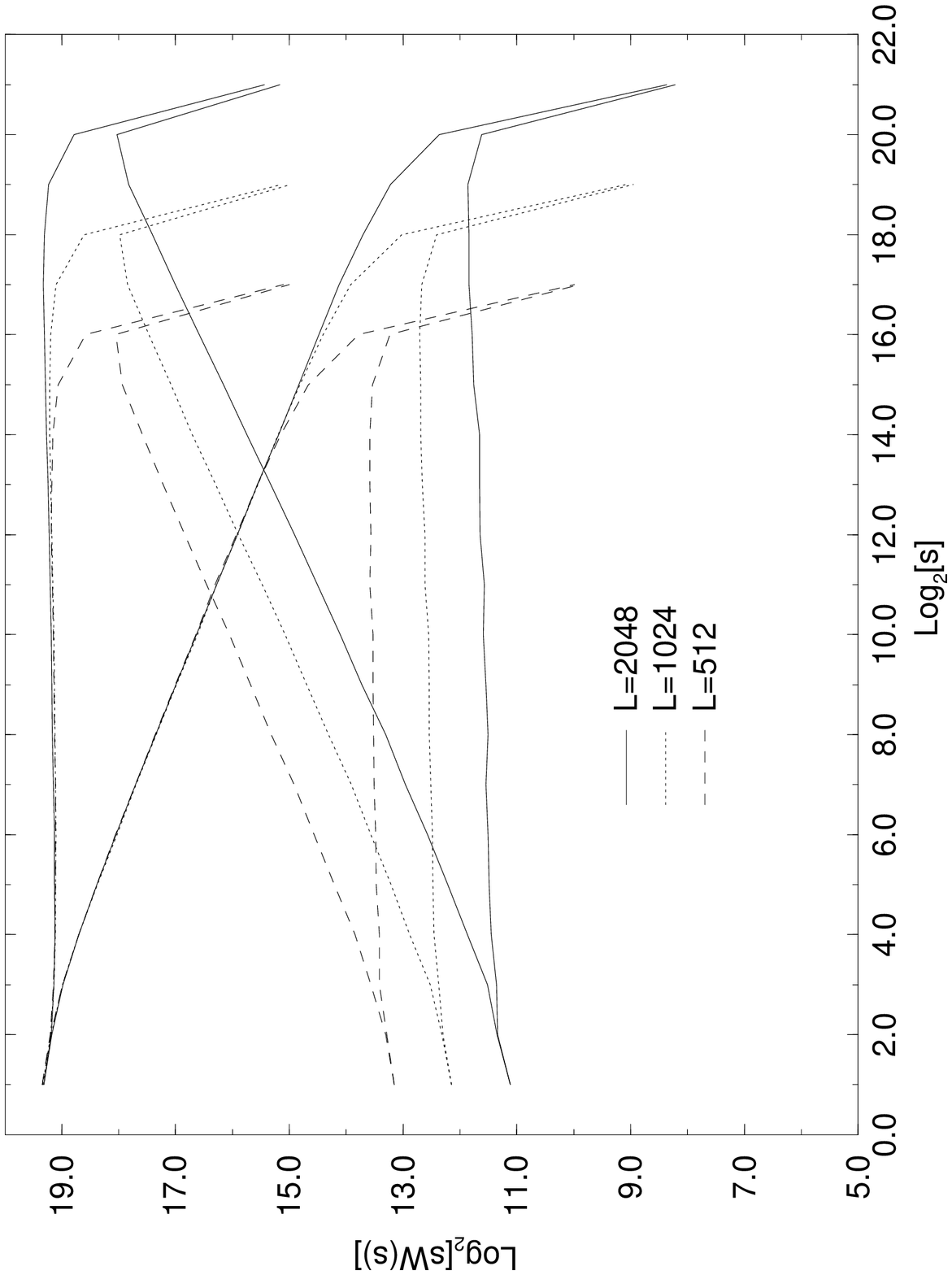}
\caption{\label{fig3}
This picture shows the distribution of all (solid line), last
(dashed line), boundary (dotted line) and last waves that touch
the boundary (dash dotted line).}
\end{figure}

The results of our numerical simulations of these
probability distributions on the lattices $L=512,1024,2048$
are shown on Fig.\ref{fig3}. They
confirm the idea of inversion symmetry with quite good accuracy.
Namely, the slopes of the linear parts of the curves
coincide with our theoretical predictions within the error
bars $\pm 0.02$. To verify our predictions about the
dependence of the probability distributions on the system
size, $L$, we studied finite-size scaling
with the universal scaling functions
\begin{mathletters}
\begin{eqnarray}
W_{all}(s)~{\rm d}s&\sim&
f_{1}\left(\frac{s}{L^2}\right)\frac{{\rm d}s}{s},\\
W_{diss}(s)~{\rm d}s&\sim&
f_{2}\left(\frac{s}{L^2}\right)\frac{{\rm d}s}{s},\\
W_{last}(s)~{\rm d}s&\sim&
L^{-3/4}f_{3}\left(\frac{s}{L^2}\right)\frac{{\rm d}s}{s},\\
W_{diss~\!\!\&~\!\!last}(s)~{\rm d}s&\sim&
L^{-3/4}f_{4}\left(\frac{s}{L^2}\right)\frac{{\rm d}s}{s}.
\end{eqnarray}
\end{mathletters}
The collapse of our data for different lattices $L=512,1024,2048$
is shown on Fig.4.

\begin{figure}
\epsfxsize=70mm
\vbox to2.9in{\rule{0pt}{2.9in}}
\includegraphics{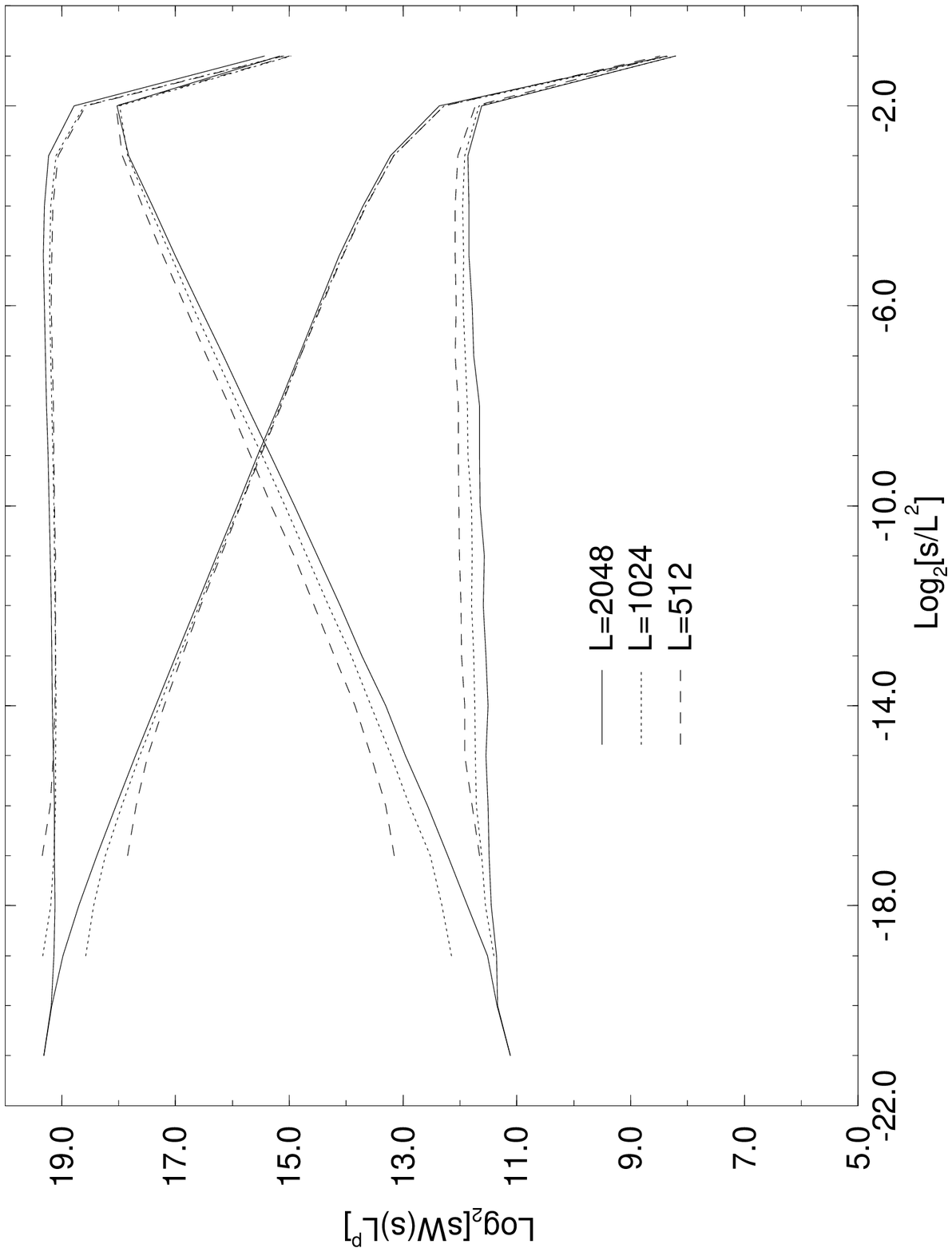}
\caption{\label{fig4}
This picture shows the collapse of our data in the finite-size analysis.}
\end{figure}

Up to now we have only been concerned about one-wave
probability distributions. These obey relatively simple
scaling laws since they
reflect rather static than dynamical properties
of the sandpile model.  The very first step
towards understanding of quite nontrivial dynamics of the
sandpile model \cite{NumSym,demenech-98} is the analysis of
two-wave distributions. This has recently been initiated by
Paczuski and Boettcher \cite{paczuski-97} who studied
numerically the transition probability distribution of the
size of the next wave $P(s_{t+1}|s_t)$, given the size of
the preceding wave $s_t$.  They suggested the following
scaling form to describe their numerical data
$P(s_{t+1}|s_{t})\sim s_{t+1}^{-\beta}~f(s_{t+1}/s_{t}),$
with the universal scaling function $f(x)$ such that
$f(x\to 0)=1$ and $f(x\gg 1)\sim x^{-r}$ and scaling
exponents $\beta\approx 0.75$ and $r\approx 0.5$.  They,
however, noticed that this scaling function pull together
the curves only for the one tail of the probability
distribution. The other tail remains loose. Hence, the very
existence of such a universal scaling function remains
dubious.

\begin{figure}
\epsfxsize=70mm
\vbox to2.9in{\rule{0pt}{2.9in}}
\includegraphics{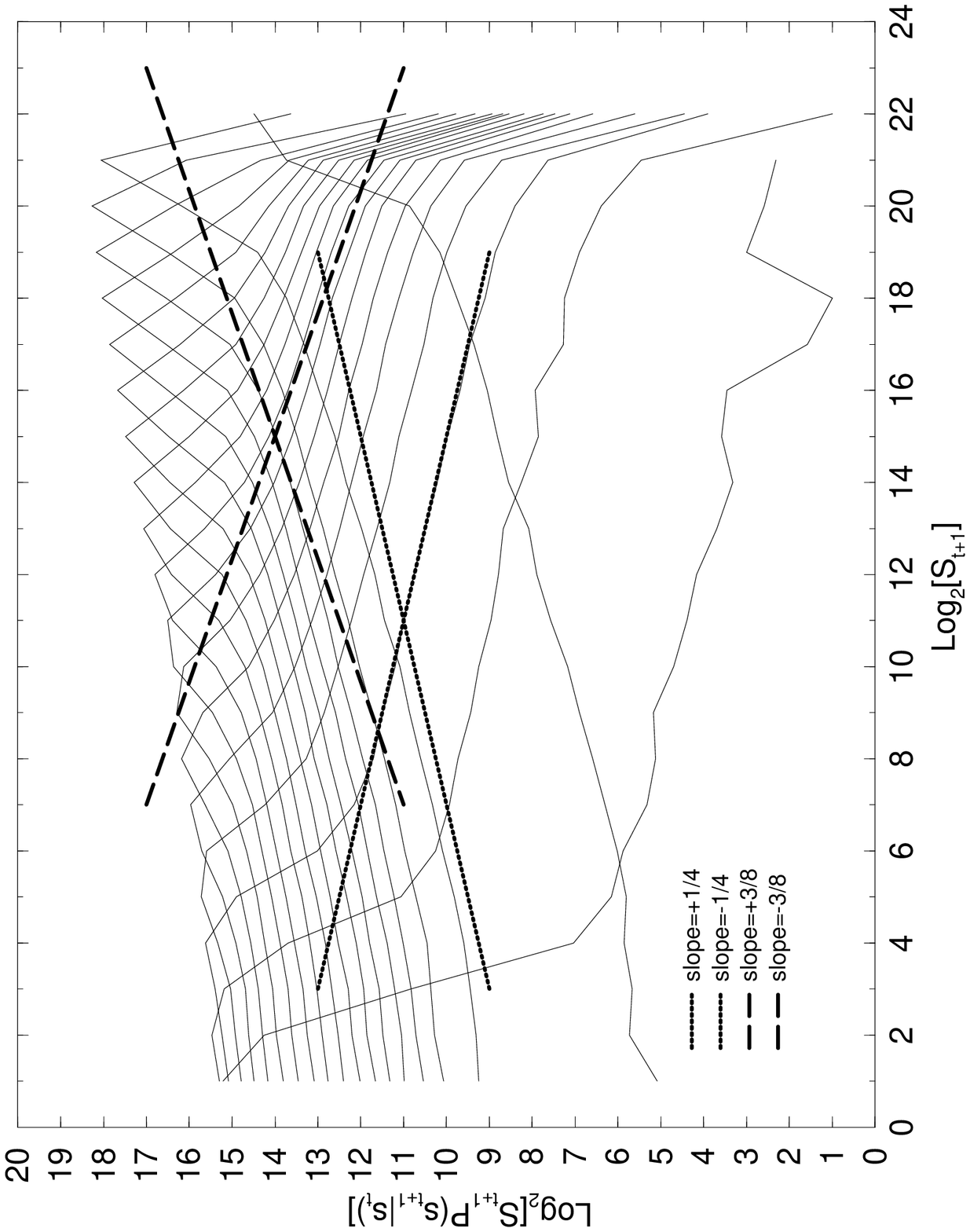}
\caption{\label{fig5}
Forward conditional probability distribution $P(s_{t+1}|s_{t})$.}
\end{figure}

\begin{figure}
\epsfxsize=70mm
\vbox to2.9in{\rule{0pt}{2.9in}}
\includegraphics{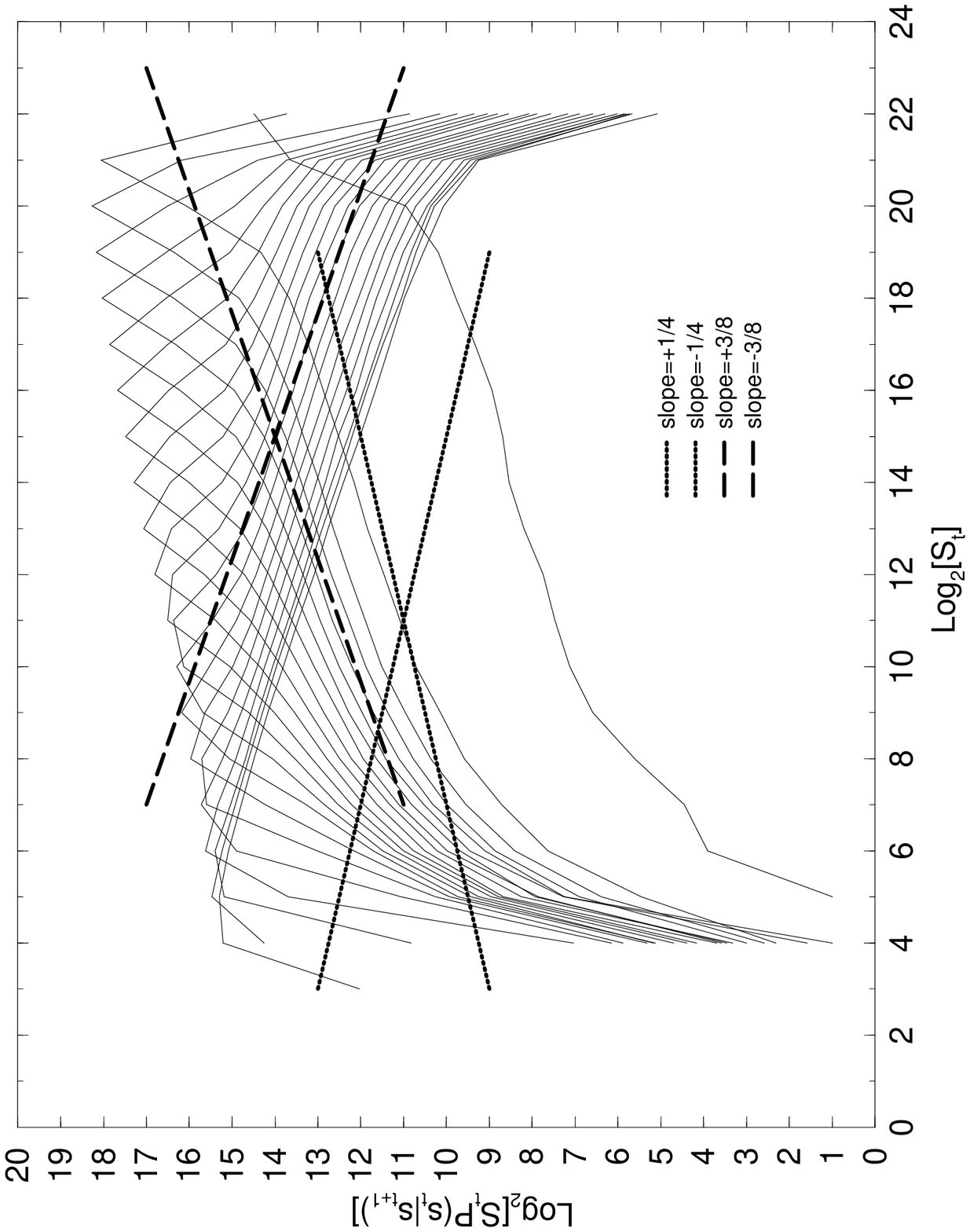}
\caption{\label{fig6}
Backward conditional probability distribution $P(s_{t}|s_{t+1})$.}
\end{figure}

The existence of
the scaling function would actually
mean that wave random process is symmetric under time
reversion. To clarify the issue we performed extensive
numerical simulations on the lattice $L=2048$.  We studied
both forward, $P(s_{t+1}|s_{t})$, and backward,
$P(s_{t}|s_{t+1})$, probability distributions. Our results
are shown on Figs.\ref{fig5} and \ref{fig6}.
We found that asymptotics of forward probability distribution do obey
scaling laws
\begin{equation}
P(s_{t+1}|s_t)\sim
        \left\{
\begin{minipage}{4.5cm}
$s_{t+1}^{-1+\gamma_1}$,~~{\rm if}~~$s_{t+1}\ll s_t$ \\
$s_{t+1}^{-1-\gamma_2}$,~~{\rm if}~~$s_{t+1}\gg s_t$
\end{minipage}
        \right.
\label{forward}
\end{equation}
with scaling exponent $\gamma_1\approx\gamma_2\approx 1/4$
proposed by Paczuski and Boettcher \cite{paczuski-97}.
However, for backward in time probability distribution we
found different asymptotics
\begin{equation}
P(s_{t}|s_{t+1})\sim
        \left\{
\begin{minipage}{4.5cm}
$s_{t}^{-1+\delta_1}$,~~{\rm if}~~$s_{t}\ll s_{t+1}$ \\
$s_{t}^{-1-\delta_2}$,~~{\rm if}~~$s_{t}\gg s_{t+1}$
\end{minipage}
        \right.
\label{backward}
\end{equation}
with scaling exponent $\delta_1\approx\delta_2\approx 3/8$.
Hence, according to our numerical simulations the wave
process is not symmetrical under
"time" reversion.

The explanation of this asymmetry is still the problem for
the future. Scaling arguments, however, can be given in favor
of the hypothesis that the exponent 3/8 is exact. First of all,
in the limit $s_{t+1}\to a^2$ the backward conditional
probability $P(s_{t}|s_{t+1})$ should coincide with the
probability distribution of the very last waves in avalanches.
Indeed, in this limit every wave of topplings which is much smaller
then the preceding one necessarily touches somewhere the boundary
of the preceding wave. Now, if we coarse grain the picture in such
a way that the size of the next wave $s_{t+1}$ becomes comparable with
the unit cell of new coarse grained lattice, then the geometry of the
picture will resemble that of last waves of topplings with the wave
$s_{t+1}$ playing the role of the origin of the avalanche on that
scale. Hence, the factor $(a^2/s)^{3/8}$ in Eq.(2) corresponds to the
factor $(s_{t+1}/s_t)^{3/8}$ in the distribution
\begin{equation}
P(s_t|s_{t+1}) \sim \left( \frac{s_{t+1}}{s_t} \right)^{3/8}
\frac{{\rm d} s_t}{s_{t}}
\end{equation}
in the limit $s_t \gg s_{t+1}$
From this it follows that exponent $\delta_2=3/8$ should be exact.
Also, if we assume that the inversion symmetry is valid not only for
one-wave but also for two-wave distributions we
get scaling relations $\gamma_1=\gamma_2$ and $\delta_1=\delta_2$.

This work was supported by the Russian Foundation for Basic
Research through Grants No. 99-01-00882 and National Science
Council of the Republic of China (Taiwan) under grant No.
NSC 89-2112-M001-005.


\begin{thebibliography}{99}
\bibitem{bak-87} P.~Bak, C.~Tang, and K.~Wiesenfeld,
  Phys. Rev. Lett. {\bf 59} (1987) 381.
\bibitem{kolmogorov-41} A.N.~Kolmogorov,
  C. R. Acad. Sci. USSR {\bf 30} (1941) 301.
\bibitem{kardar-86} M.~Kardar, G.~Parisi and Y.-C.~Zhang,
  Phys. Rev. Lett. {\bf 56},889 (1986).
\bibitem{NumSym}
  L.P.~Kadanoff, S.R.~Nagel, L.~Wu and S.M.~Zhou,
  Phys. Rev. A {\bf 39} (1989) 6524;
  S.S.~Manna, J. Stat. Phys. {\bf 59} (1990) 509;
  P.~Grassberger and S.S.~Manna, J. Phys. France
  {\bf 51} (1990) 1077.
\bibitem{dhar-90} D.~Dhar,
  Phys. Rev. Lett. {\bf 64} (1990) 1613.
\bibitem{majumdar-92} S.N.~Majumdar and D.~Dhar,
  Physica A {\bf 185} (1992) 129.
\bibitem{majumdar-91} S.N.~Majumdar and D.~Dhar,
  J. Phys. A {\bf 24} (1991) L357.
\bibitem{priezzhev-94} V.B.~Priezzhev,
  J. Stat. Phys. {\bf 74} (1994) 955.
\bibitem{ivashkev-94a} E.V.~Ivashkevich,
  J. Phys. A {\bf 27} (1994) 3643.
\bibitem{demenech-98}  M.~De~Menech, A.L.~Stella and C.~Tebaldi,
  Phys. Rev. E {\bf 58} (1998) R2677;
\bibitem{ivashkev-94b} E.V.~Ivashkevich, D.V.~Ktitarev and V.B.~Priezzhev,
  Physica A {\bf 209} (1994) 347.
\bibitem{dhar-94} D.~Dhar and S.S.~Manna,
  Phys. Rev. E {\bf 49} (1994) 2684.
\bibitem{ivashkev-94c} E.V.~Ivashkevich, D.V.~Ktitarev and V.B.~Priezzhev,
  J. Phys. A  {\bf 27} (1994) L585.
\bibitem{priezzhev-96} V.B.~Priezzhev, D.V.~Ktitarev and E.V.~Ivashkevich,
  Phys. Rev. Lett. {\bf 76} (1996) 2093.
\bibitem{paczuski-97} M.~Paszuski and S.~Boettcher,
  Phys. Rev. E {\bf 56} (1997) R3745.
\bibitem{saleur-87} H.~Saleur and B.~Duplantier,
  Phys.Rev.Lett. {\bf 58} (1987) 2325.
\end{thebibliography}
\end{document}